\documentclass[aps,pra,floats,floatfix,twocolumn,longbibliography,superscriptaddress]{revtex4-1}

\usepackage{color} 
\usepackage{latexsym}
\usepackage{amsmath} 
\usepackage{amssymb} 
\usepackage{bm}
\usepackage{wasysym}

\usepackage{graphicx}
\usepackage{epstopdf}
\graphicspath{{./}{./Figs/}}

\usepackage[colorlinks,linkcolor=blue,urlcolor=blue,bookmarks=true,pdftitle={}]{hyperref}

\newcommand{\beq}{\begin{eqnarray}}
\newcommand{\eeq}{\end{eqnarray}} 

\newcommand{\hide}[1]{}
\newcommand{\Eq}[1]{\textcolor{blue}{Eq.\!\!~(\ref{#1})}} 
\newcommand{\Fig}[1]{\textcolor{blue}{Fig.}\!\!~\ref{#1}}

\begin{document}

\title{Chaos and bi-partite entanglement between Bose-Joephson junctions}

\author{Amichay Vardi} 
\affiliation{Department of Chemistry, Ben-Gurion University of the Negev, Beer-Sheva 84105, Israel}

\begin{abstract}
The entanglement between two weakly coupled bosonic Josephson junctions is studied in relation to the classical mixed phasespace structure of the system, containing symmetry-related regular islands separated by chaos. The symmetry-resolved entanglement spectrum and bi-partite entanglement entropy of the system's energy eigenstates are calculated and compared to their expected structure for random states that exhibit complete or partial ergodicity. The entanglement spectra of chaos-supported eigenstates match the  microcanonical structure of a Generalized Gibbs Ensemble due to the existence of an adiabatic invariant that restricts ergodization on the energy shell. The symmetry-resolved entanglement entropy of these quasistochastic states consists of a mean-field maximum entanglement term and a fluctuation correction due to the finite size of the constituent subsystems. The total bi-partite entanglement entropy of the eigenstates correlates with their chaoticity.  Island-supported eigenstates are macroscopic Schr\"odinger cat states for particles and excitations, with substantially lower entanglement.
\end{abstract}
\maketitle


\section{Introduction}
%
The study of quantum entanglement has lately focused on many-body systems, with important applications in quantum information and condensed matter physics \cite{Amico08,Laflorencie16}. Entanglement lies at the heart of quantum information processing \cite{Deutch85} and quantum teleporatation \cite{Plenio98}. In condensed matter physics, it underlies the density matrix renormalization group methodology \cite{White92,Schollwock05,Schollwock11}, quantum phase transitions \cite{Osterloh02,Osborne02,Gu04} and topological order \cite{Li08}, quantum quench dynamics \cite{Calabrese05,Eisler07,Calabrese07,Santos11}, quantum thermalization \cite{Tikhonenkov13,Khripkov15,Kauffman16,Khripkov20,Brenes20}, and many-body localization \cite{Kjall13,Huang14}.

Considerable effort has recently been concentrated towards the study of bi-partite entanglement in the stochastic-like eigenstates of quantum chaotic Hamiltonians \cite{Page93,Deutch10,Santos12,Hamma12,Znidaric07,Alba15,Beugling15,Vidmar17,Garrison18,Murthy19,Huang19,Lydzba20,Lydzba21,Haque22}. The entanglement entropy of such states is near maximal, because chaotic ergodization implies the eigenvalues of the reduced  subsystem density matrices are spread nearly uniformly, with a fluctuation correction due to the finite size of the subsystems \cite{Page93,Vidmar17}.

So far, the analysis of eigenstate entanglement has focused on systems that exhibit 'hard' chaos, i.e. their entire phasespace is chaotic. However, in many-body systems with few classical degrees of freedom the phase space is often {\em mixed}, with 'islands' of quasi-integrability due to the local conservation of residual motional constants, interspersed between chaotic 'seas' in which the dynamics is ergodic. The ergodicity of such systems is incomplete and corresponds to {\em generalized Gibbs ensembles} (GGEs) rather than the canonical ensembles encountered in the presence of relaxation and pumping, or the microcanonical ensembles obtained for isolated systems. It is thus desirable to establish how bi-partite quantum entanglement is affected by the partition of the classical phasespace into regular and chaotic regions.

In this work, the $U(1)$ symmetry resolved entanglement spectrum and entanglement entropy are studied for the minimal model system of two weakly-coupled bosonic Josephson junctions. In a sense, this is the interacting many-body bosonic equivalent of the ubiquitous two qubits system in which the notion of bi-partite entanglement first appears \cite{Avron09}. It was previously shown \cite{Strzys10,Strzys12} that the timescale separation between the fast internal motion within each junction and the slow exchange of particles and energy between them, implies the adiabatic invariance of the total number of Josephson excitations $J$ corresponding in the classical limit to the sum of subsystem actions,  in addition to the obvious conservation of the total system energy $E$ and the number of particles $N$. The adiabatic system dynamics can thus be described as the slow motion of particles and Josephson quasiparticles ('josons') between the constituent subsystems. 

While for small perturbations around the stationary points the classical dynamics of this system reduces to coupled Josephson oscillations \cite{Strzys10,Strzys12,Khripkov13,Khripkov14}, at higher energy it is surprisingly rich. In particular,  we find that the mutual conservation of $E$ and $J$ generates a mixed phase space structure, with integrable self-trapping islands of two types separated by a chaotic sea. The quantum eigenstates are correspondingly supported by the different classical phase-space regions. The bi-partite entanglement entropies of the system's eigenstates are correlated with ergodicity measures such as the participation number and the Shanon entropy. The chaos-supported eigenstates exhibit the expected near maximal entanglement. By contrast, island-supported eigenstates are macroscopic Schr\"odinger cat states involving only ${\cal O}(2)$ nonvanishing eigenvalues in the reduced subsystem density matrices. The population imbalance distribution and the symmetry resolved entanglement entropy of the chaotic eigenstates corresponds to a GGE that accounts for the adiabatic invariance of $J$ within the respective energy shell.

The model system is introduced in Sect.~\ref{model}, its adiabatic dynamics is discussed in Sect.~\ref{adiabatic}, and the methodology for evaluation the number of excitations is briefly recalled in Sect.~\ref{jos}. The system's eigenstates and their relation to the mixed classical phasespace are discussed in Sect.~\ref{eigenstates}. The bi-partition of the system and the  reduced subsystem density matrices are defined in Sect.~\ref{redrho} and the symmetry-resolved entanglement spectrum is presented in Sect.~\ref{nrent}. The expected particle imbalance and bi-partite entanglement entropy distributions for ergodic and semiergodic states are compared in Sect.~\ref{ergodic} with the numerically calculated distributions of the chaos-supported eigenstates, demonstrating the agreement with the GGE prediction. Summary and concluding remarks are provided in Sect.~\ref{conclusions}.


\section{The coupled dimers model}
\label{model}
Consider a system of two weakly-coupled bosonic Josephson junctions (a.k.a 'Bose-Hubbard dimers'), described by the four-mode Hamiltonian \cite{Strzys10,Strzys12,Khripkov13,Khripkov14}
\beq
\hat{H}&=& -\frac{\Omega}{2}\left( \sum_\alpha \hat{a}_{+,\alpha}^\dag\hat{a}_{-,\alpha} + H.c\right)+ \frac{U}{2}\sum_{\alpha,\sigma}\hat n_{\sigma,\alpha}\left(\hat{n} _{\sigma,\alpha}-1\right)\nonumber \\
&-& \frac{\omega}{2} \left(\sum_\sigma \hat{a}^\dag_{\sigma,{\rm L}}\hat{a}_{\sigma,{\rm R}}+\hat{a}^\dag_{\sigma,{\rm R}}\hat{a}_{\sigma,{\rm L}}\right)
\label{Dicke_ham}
\eeq
where, $\hat a_{\sigma,\alpha}$ annihilate a boson in the $\sigma=\pm$ mode of the $\alpha={\rm L},{\rm R}$ junction. The inter-dimer coupling $\omega$ is assumed to be much smaller than the coupling $\Omega$ between the two modes of each dimer and the on-site interaction strength $U$. Below we rescale time as $t\rightarrow \Omega t$ so that frequencies are given in units of $\Omega$ and the dimensionless system parameters are $w=\omega/\Omega\ll1$ and $u=UN/\Omega$. 

Accounting for the conservation of the total number of particles $N$, the Hilbert space dimension of the many-body system is $D=(N+1)(N+2)(N+3)/6$. In the limit of large $N$, the restricted coherent-state (mean field) dynamics is obtained by replacing the operators $\hat a_{\sigma,\alpha}$ by $c$-numbers. The resulting classical motion has $d=3$ degrees of freedom, e.g. three population imbalances and three relative phases between the classical amplitudes serving as action-angle variables.

\section{Adiabatic dynamics}
\label{adiabatic}
The dynamics of the double-dimer model in the adiabatic limit $w \ll 1$ was reduced by Strzys and Anglin \cite{Strzys10,Strzys12}, to the slow exchange of particles and  `josons' between the two subsystems. Their procedure begins with a Holstein-Primakoff transformation (HPT) applied to Eq.~(\ref{Dicke_ham}):
\begin{eqnarray}
\label{HPT}
\frac{n_{\alpha}}{2}-\hat{\mathcal{A}}_\alpha ^{\dagger} \hat{\mathcal{A}}_\alpha &\equiv & \frac{1}{2}(\hat{a}_{\alpha,+}^{\dagger}\hat{a}_{\alpha,-}+\hat{a}_{\alpha,-}^{\dagger}\hat{a}_{\alpha,+})~, \\
\sqrt{n_{\alpha}-\hat{\mathcal{A}}_\alpha ^{\dagger} \hat{\mathcal{A}}_\alpha} \hat{\mathcal{A}}_\alpha &\equiv & \frac{1}{2}(\hat{a}_{\alpha,+}^{\dagger}+\hat{a}_{\alpha,-}^{\dagger})(\hat{a}_{\alpha,+}-\hat{a}_{\alpha,-})~, \nonumber\\
\hat{\mathcal{A}}_\alpha^{\dagger} \sqrt{n_{\alpha}-\hat{\mathcal{A}}_\alpha ^{\dagger} \hat{\mathcal{A}}_\alpha} &\equiv & \frac{1}{2}(\hat{a}_{\alpha,+}^{\dagger}-\hat{a}_{\alpha,-}^{\dagger})(\hat{a}_{\alpha,+}+\hat{a}_{\alpha,-})~.\nonumber
\end{eqnarray}  
The operators $\hat{\mathcal{A}}_\alpha$ shift atoms between the two $\sigma=\pm$ modes of the $\alpha=\rm \{L,R\}$ dimer.  They obey the commutation relation, $[\hat{\mathcal{A}}_\alpha, \hat{\mathcal{A}}_\alpha^{\dagger}]=1$ and $[\hat{\mathcal{A}}_\alpha, \hat{n}_\alpha]=0$.  Consequent application of the Bogoliubov transformation, $\hat{\mathcal{A}}_\alpha=u_\alpha \hat{\mathcal{J}}_\alpha+v_\alpha \hat{\mathcal{J}}_\alpha^{\dagger}$, transform the single dimer Hamiltonian as,
\begin{eqnarray}
\hat{H}_{\alpha} &=& -\frac{\Omega}{2} (\hat{a}_{+,\alpha}^\dag\hat{a}_{-,\alpha} + {\rm h.c.})+ \frac{U}{2}\sum_{\sigma}\hat n_{\sigma,\alpha}\left(\hat{n} _{\sigma,\alpha}-1\right) \nonumber \\
&\rightarrow & \frac{\Omega}{2}\hat{n}_\alpha + \frac{U}{4}\hat{n}_\alpha (\hat{n}_\alpha-2)+ \sqrt{\Omega(\Omega +U\hat{n}_\alpha)}\hat{\mathcal{J}}_\alpha^{\dagger}\hat{\mathcal{J}}_\alpha\nonumber\\
&~& - \frac{U}{8}\frac{4\Omega+U\hat{n}_\alpha}{\Omega+U\hat{n}_\alpha} \hat{\mathcal{J}}_\alpha^{\dagger 2}\hat{\mathcal{J}}_\alpha^2 + \mathcal{O}(Un_{\alpha}^{-1})
\label{HPT_ham}
\end{eqnarray}
where, $u_\alpha$ and $v_\alpha$ are quasi-hole and particle excitation amplitudes, respectively, and $\hat{\mathcal{J}}_\alpha$ obeys the bose commutation relation, $[\hat{\mathcal{J}}_\alpha, \hat{\mathcal{J}}_\alpha^{\dagger}]=1$. In deriving Eq.~\ref{HPT_ham} terms that do not commute with $\hat{\mathcal{J}}_\alpha^{\dagger}\hat{\mathcal{J}}_\alpha$ have been neglected. A second HPT applied to the inter-dimer hopping, reads in the large-N limit,
\begin{equation}
\hat{n}_{\rm L,R}=\frac{1}{2}[N\pm N^{1/2}(\hat{\mathcal{A}}^{\dagger}+\hat{\mathcal{A}})]
\label{HPT1}
\end{equation}  
where, $\hat{\mathcal{A}}$ shifts atoms between the junctions, and obeys the commutation relation, $[\hat{\mathcal{A}},\hat{\mathcal{A}}^{\dagger}]=1$.  Equation~(\ref{HPT1}) retains total number conservation. Hence , the total Hamiltonian in \Eq{Dicke_ham}, including the single dimer hamiltonians $\hat{H}_\alpha$ and the interdimer coupling, can be written (in units of $\Omega$) as,
\begin{eqnarray}
\label{joson_ham}
\hat{H} &\rightarrow& w \hat{\mathcal{A}}^{\dagger}\hat{\mathcal{A}} + 
\frac{u}{8}(\hat{\mathcal{A}}^{\dagger}+\hat{\mathcal{A}})^2 \\
~&~&- \frac{w_{\mathcal{J}}}{2}(\hat{\mathcal{J}}_{\rm L}^{\dagger}\hat{\mathcal{J}}_{\rm R} + \hat{\mathcal{J}}_{\rm R}^{\dagger}\hat{\mathcal{J}}_{\rm L})- \frac{U_{\mathcal{J}}}{2}\sum_{\alpha=\rm L,R}\hat{\mathcal{J}}_{\alpha}^{\dagger 2}\hat{\mathcal{J}}_{\alpha}^2  \nonumber\\
~&~&+ \frac{u}{4}\sqrt{\frac{1}{1+u/2}}\frac{(\hat{\mathcal{A}}+\hat{\mathcal{A}}^{\dagger})}{\sqrt{N}}(\hat{\mathcal{J}}_{\rm L}^{\dagger}\hat{\mathcal{J}}_{\rm L}-\hat{\mathcal{J}}_{\rm R}^{\dagger}\hat{\mathcal{J}}_{\rm R})\nonumber
\end{eqnarray}
where, the effective tunnelling frequency and interaction strength of the Josephson excitations are given by,
\begin{equation}
w_{\mathcal{J}} = w \frac{1 + u/4}{\sqrt{1+u/2}} \quad \text{and} \quad U_{\mathcal{J}} = U\frac{1+u/8}{1+u/2},
\end{equation}
respectively. The two first terms on the r.h.s. of Eq.~(\ref{joson_ham}) correspond to Josephson oscillations of particles whereas the third and fourth terms are a Josephson Hamiltonian for the excitations with effective attractive interaction between them. The last term couples the two oscillations (due to the dependence of the fast internal dimer frequencies on particle number). In addition to $N$, the total number of excitations $J=\sum_{\alpha={\rm L,R} }{\mathcal J}_\alpha^\dag {\mathcal J}_\alpha\rightarrow j_L+j_R$ is also conserved by the approximate Hamiltonian~(\ref{joson_ham}). Thus, while the conservation of $N$ is strict, $J$ is an adiabatic invariant.

\begin{figure}
\centering
\includegraphics[clip=true, width=\columnwidth]{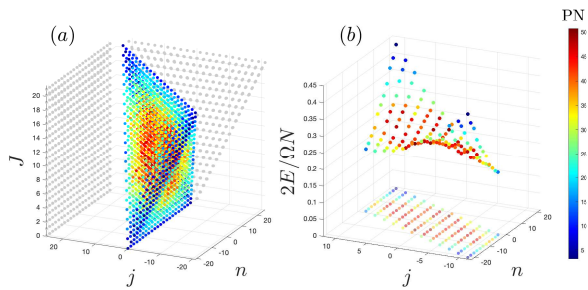}
\caption{{\bf  Spectrum of decoupled dimers:} (a) The eigenstates of the double-dimer model with $N=21, \omega=0, u=0.5$, arranged according to the good quantum numbers  $n$,$j$, and $J$. Each eigenstate is colored according to its participation number in the exact eigenbasis with the same parameters except $\omega=0.082$. Grey points mark the projections onto the $\{n,J\}$ and $\{j,J\}$ planes; (b) The energies of the $J=11$ shell of the spectrum in (a) and its projection onto the $\{n,j\}$ plane.}
\label{UnpertSpec}
\end{figure}

\section{Number of excitations}
\label{jos}
The number of excitations in  the two subsystems was evaluated in Ref.~\cite{Ray20}, using a semiclassical approach. Given the energy $E_\alpha$ and the number of particles $n_\alpha$ of the $\alpha={\rm L,R}$ Bose-Josephson subsystem, the number of excitations $j_\alpha$ is just the classical action, i.e. the phase space area enclosed by the classical energy contour in units of the effective Planck constant $h=4\pi/n_\alpha$. This area can be analytically evaluated for the elliptical energy contours encountered as long as $u_\alpha\equiv Un_\alpha/\Omega\leq 1$, resulting in the expression,
\begin{eqnarray}
j_\alpha& =&\frac{1}{2} - \frac{1}{\pi}\mathrm{Re}\left[ 2\cos\eta_\alpha\frac{K\!\!\left(\frac{1-i e^{-i \eta_\alpha } u_\alpha}{1+ie^{i \eta_\alpha }u_\alpha}\right)}{\sqrt{1+i e^{i \eta_\alpha } u_\alpha}}\right.\nonumber\\
~&~&+\frac{2i}{u_\alpha}\sqrt{1+i e^{i \eta_\alpha } u_\alpha}\, E\!\!\left(\frac{1-i e^{-i \eta_\alpha} u_\alpha}{1+ie^{i \eta_\alpha }u_\alpha}\right)\\
~&~&\left.+\frac{i e^{i \eta_\alpha } \left(1-i e^{-i \eta_\alpha } u_\alpha\right) \Pi\!\!\left(i e^{-i \eta_\alpha } u_\alpha|\frac{2 i u_\alpha \cos  \eta_\alpha }{1+ie^{i \eta_\alpha } u_\alpha}\right)}{\sqrt{1+i e^{i \eta_\alpha } u_\alpha}}\right],\nonumber
 \label{jdef}
\end{eqnarray}
where $K(m)$, $E(m)$, and $\Pi(n|m)$ are respectively the complete elliptic integrals of the 1st, 2nd and 3rd kinds, and $\eta_\alpha=\arcsin[2E_\alpha/(\Omega n_\alpha)]$ is used to parametrize $-\Omega n_\alpha/2<E_\alpha<\Omega n_\alpha/2$ to the $[-\pi/2,\pi/2]$ range.

Below we denote the the total particle and excitation numbers as $N=n_{\rm L}+n_{\rm R}$ and $J=j_{\rm L}+j_{\rm R}=0,1,...,N$, respectively, and the corresponding particle and excitation imbalances as $n=n_{\rm L}-n_{\rm R}=-N,...,N$ and $j=j_{\rm L}-j_{\rm R}=-J,...J$, respectively.

\section{Eigenstates}
\label{eigenstates}
\subsection{The unperturbed basis}
In the absence of interdimer coupling ($\omega=0$), the two-dimer energy eigenstates are direct products of single dimer eigenstates in the form,
\beq
|\mu\rangle=|N,J,n,j\rangle=| n_{\rm L}, j_{\rm L}\rangle\otimes| n_{\rm R}, j_{\rm R}\rangle
\label{upstates}
\eeq
with $N,J,n,j$ being good quantum numbers, as illustrated in \Fig{UnpertSpec}a. The dimension of each fixed-$J$ shell in the spectrum is $D^{(J)}=(J+1)(N+1-J)$ and the number of states within the shell with a given $n_{\rm L}$ is,
\beq
D^{(J,n_{\rm L})}=\left\{\begin{array}{lc}
n_{\rm L}+1&0\leq n_{\rm L}\le J\\
J+1&J<n_{\rm L}<N-J\\
N-n_{\rm L}+1&N-J\leq n_{\rm L}\leq N
\end{array}\right.
\eeq
The energies of a single $J$ shell in the middle of the spectrum are plotted in \Fig{UnpertSpec}b. Within this shell, the energies of high $|n|$ eigenstates are elevated due to the repulsive interaction between particles. By contrast, the energies of high $|j|$ eigenstates are lowered due to the effective repulsion between the Josephson excitations in \Eq{joson_ham}.

\begin{figure}
\centering
\includegraphics[clip=true, width=\columnwidth]{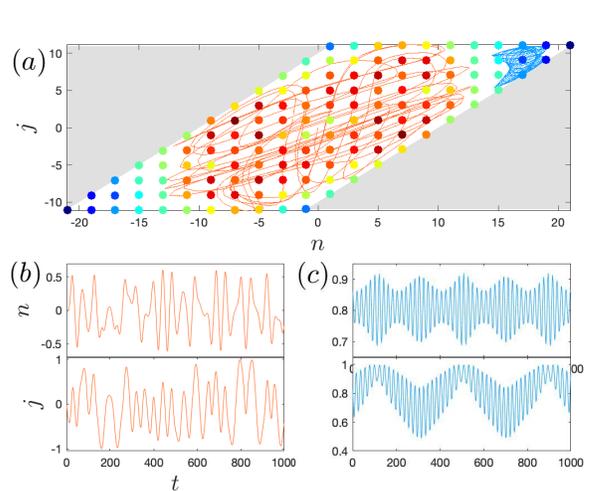}
\caption{{\bf Mixed structure of the fixed $J$ shells:} (a) The $J=11$ shell of the coupling-free spectrum from \Fig{UnpertSpec} overlaid on representative classical trajectories launched at $J=11$, $n=j=0$ (orange) and $J=11$, $n=18.2$, $j=9.3$ (blue), with the same parameters. The relative interdimer phase of both trajectories is $\varphi_{\rm LR}=0$. The dynamics of the particle and joson imbalances $n,j$ for the same trajectories is plotted in panels (b) and (c), respectively.}
\label{QCC}
\end{figure}

\subsection{Exact spectrum}
For finite $\omega$ we can numerically diagonalize the Hamiltonian in \Eq{Dicke_ham} to obtain the exact eigenstates $|\nu\rangle$. Projecting the unperturbed states of \Eq{upstates} onto the exact basis to obtain $p_{\nu,\mu}=|\langle \nu|\mu\rangle|^2$, we can calculate the {\em participation number} ${\rm PN}_\mu=\left(\sum_\nu p_{\nu,\mu}^2\right)^{-1}$, estimating the number of exact eigenstates that contribute to the unperturbed state $|\mu\rangle$. The participation numbers in the $\omega=0.082$ basis are denoted by color in \Fig{UnpertSpec}. It is clear that the mid-spectrum fixed $J$ surfaces contain two pairs of regions with low participation, corresponding to the maxima and minima of the energy surface. These localization regions are separated by a large high-participation ergodic region around the central energy saddle point.

In \Fig{QCC} the same $J$ shell of the unperturbed spectrum is plotted over two representative classical trajectories. The participation numbers in the finite-coupling basis correlate well with the classical phasespace structure for the same parameters, which due to mutual conservation of $J$ and $E$ separates into two pairs of integrable islands in which either particles or excitations are macroscopically self-trapped, and a central chaotic region, explored ergodically by all trajectories launched in it.

\begin{figure}
\centering
\includegraphics[clip=true, width=\columnwidth]{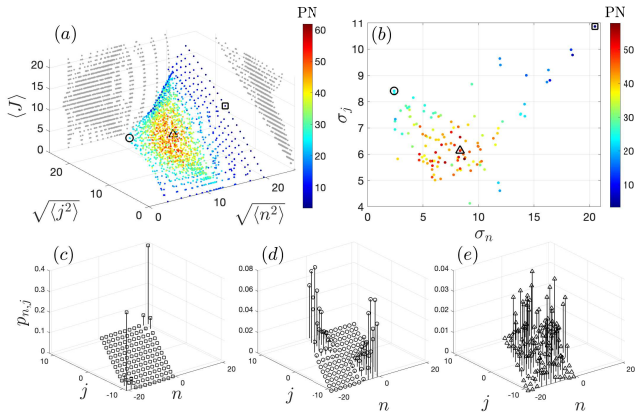}
\caption{{\bf Coupled dimers spectrum:} (a) The eigenstates of the double-dimer model with $N=21, \omega=0.082, u=0.5$, arranged according to their  particle-imbalance variance $\sigma_n=\sqrt{\langle n^2\rangle}$, excitation-imbalance variance $\sigma_j=\sqrt{\langle j^2\rangle}$, and mean number of excitations $\langle J\rangle$. Each eigenstate is colored according to its participation number in the unperturbed basis set; (b) The $J=11$ shell of the spectrum in (a). Markers in (a) and (b) mark the states with the minimum $\sigma_n$ ($\circ$), the maximum $\sigma_n$ ($\square$), and the maximum Shanon entropy $\cal H$ ($\triangle$) within this $J$ shell; (c-e) The probability distribution $p_{n,j}$ for the marked states in (a) and (b) with the same marker convention.}
\label{ExactSpec}
\end{figure}

For finite interaction, $n$ and $j$ are no longer good quantum numbers. Due to its symmetry, the exact eigenstates of the coupled-dimers system belong to one of the four irreducible representations of the dihedral group $D_2$. Therefore the expectation values of the particle and excitation imbalance are $\langle n\rangle=\langle j\rangle=0$. Therefore, in \Fig{ExactSpec}a we classify the exact eigenstates according to the standard deviations $\sigma_n=\sqrt{\langle n^2\rangle}$ and $\sigma_j=\sqrt{\langle j^2\rangle}$. Each state is colored according to its participation number in the unperturbed basis ${\rm PN}_\nu=(\sum_\mu p_{\nu,\mu}^2)^{-1}$. Since the timescale separation between fast intradimer motion and slow interdimer particle/excitation exchange is maintained, $J$ is conserved so that the exact eigenstates only mix zero-coupling states within a {\em single} $J$ shell. This is evident in the layering of the spectrum in shells with integer value of $\langle J\rangle$.

Plotting a fixed  $J=J_0$ shell of the exact spectrum in \Fig{ExactSpec}b, we identify three representative states, for which we plot the probability distribution $p_{n,j}(\nu)=|\langle \nu | N,J_0,n,j\rangle|^2$ (it was verified that $\sum_{n,j} p_{n,j}=1$, i.e. that there is no projection onto states with $J\neq J_0$). The states with high $\sigma_n$ (\Fig{ExactSpec}c) are macroscopic cat states, i.e superpositions of localized states supported by the integrable particle-self-trapping islands. Similarly, states with high $\sigma_j$ and  low $\sigma_n$ (\Fig{ExactSpec}d) are joson macroscopic cat-states, depicting similarly populated dimer subsystems with the excitations in a superposition of all-${\rm L}$ and all-${\rm R}$. In between these macroscopic superpositions that come as odd-even doublets with spacing that vanishes exponentially with $h$, lie the high participation states distributed ergodically over the classically chaotic region of the $J$-shell (\Fig{ExactSpec}e). The chaoticity of the eigenstates $|\nu\rangle$ can be quantified by their {\em Shanon entropy},
\beq
{\cal H}_\nu=-\sum_{m=1}^D p_{\nu,m}\log p_{\nu,m}\leq\log(D)={\cal H}_{\rm max}~,
\eeq
where $p_{\nu,m}=|\langle m|\nu\rangle|^2$ are the expansion probabilities of $|\nu\rangle$ in the computational Fock state basis $|m\rangle=|n_{{\rm L},+},n_{{\rm L},-},n_{{\rm R},+},n_{{\rm R},-}\rangle$. For a fully chaotic system, $p_{\nu,m}$ may be replaced by independent real random variables from a Gaussian distribution fluctuating around $1/D$, resulting in the limiting value \cite{Borgonovi16},
\beq
{\cal H}_{\rm GOE}=\log(0.48D)~.
\eeq

\begin{figure}
\centering
\includegraphics[clip=true, width=\columnwidth]{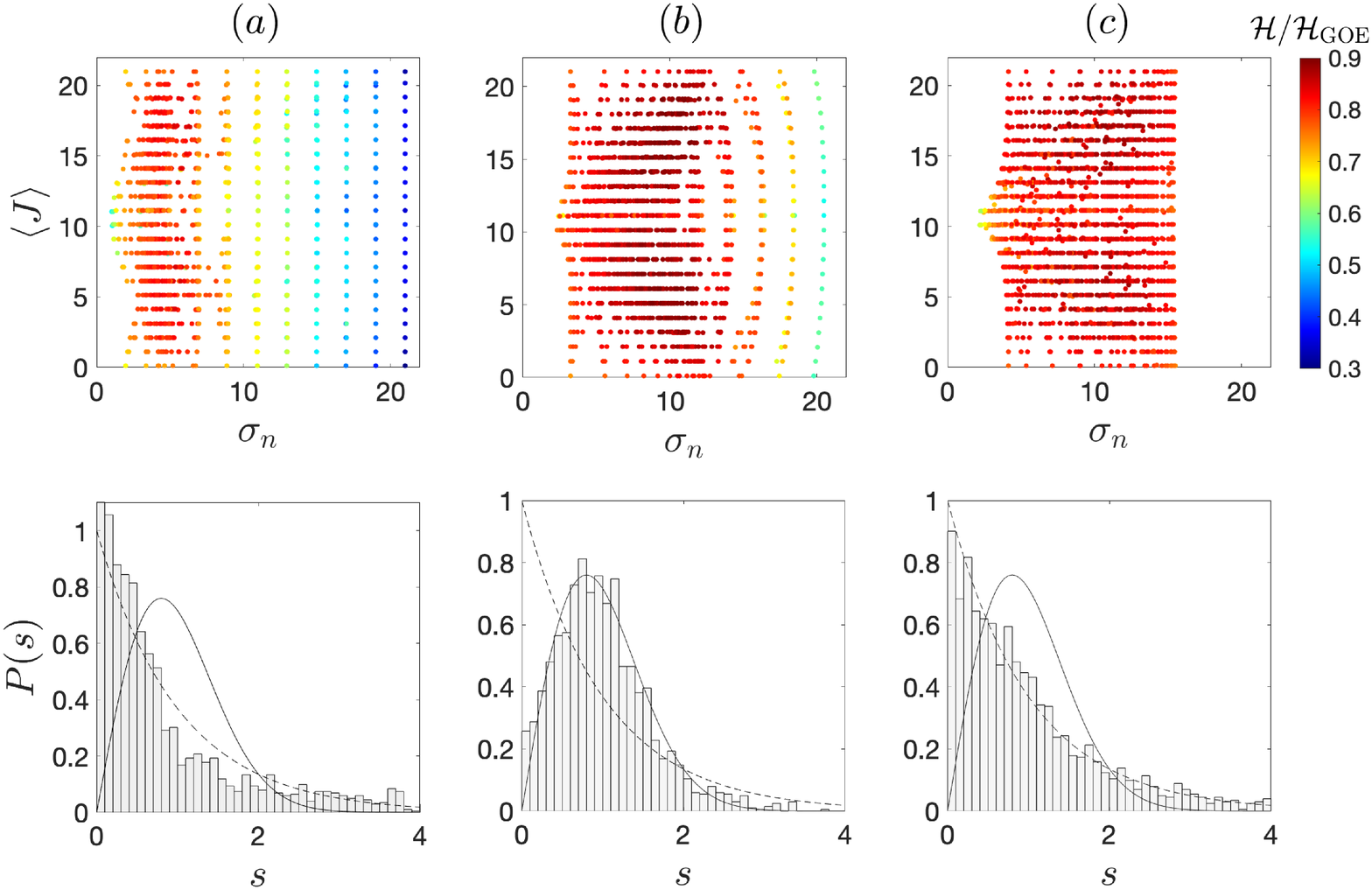}
\caption{{\bf Level spacing statistics:} The coupled-dimers spectrum (top), color-coded according to the Shanon entropy ${\cal H}$, and the level spacing statistics (bottom) at $\omega=0.01$ (a), 0.082 (b) and 0.5 (c). Other parameters are the same as in \Fig{ExactSpec}}
\label{SpSt}
\end{figure}

In \Fig{SpSt} the Shanon entropy of the coupled-dimers eigenstates is plotted for three values of the interdimer coupling $\omega$, along with the level-spacing statistics obtained by separating the spectrum to the four $D_2$ symmetry classes and unfolding each class according to the local mean spacing $\bar s(E)$. Poissonian level statistics $P(s)=e^{-s}$ indicates integrability, whereas chaos is detected by the Wigner surmise distribution $P(s)=(\pi/2)se^{-\pi s^2/4}$. When the coupling is weak (\Fig{SpSt}a) the chaotic region is small and the dynamics takes place mostly in the integrable self trapping islands, where $|n|$ is still a good quantum number. Level spacing statistics is therefore nearly Poissonian. The chaotic sea grows with $\omega$, peaking (\Fig{SpSt}b) at the value used in \Fig{UnpertSpec} and \Fig{ExactSpec} (the exact parameters for maximizing chaos were determined by analysis of the Brody parameter \cite{Brody73} and the adjacent spacing correlation function \cite{Atas13}). Further increase in $\omega$ restores integrability because the dynamics becomes linear (\Fig{SpSt}c). Note also that when $\omega$ becomes comparable with the internal dimer frequencies, joson number conservation is violated due to the breakdown of adiabaticity so that $\langle J\rangle$ can take non-integer values.

Below, we aim to characterize the number-resolved bi-partite entanglement between the Josephson qubits  for the coupled-dimers eigenstates, and correlate it with their chaos measure.

\section{Reduced subsystem density matrices}
\label{redrho}
Consider the bi-partition into the ${\rm L,R}$ dimer subsystems. The state of the system can be expanded in any arbitrary bi-partite basis,
\beq
|\psi\rangle=\sum_{n_{\rm L}=0}^N \sum_{l=1}^{d_{{\rm L}}(n_{\rm L})}\sum_{r=1}^{d_{\rm R}(n_{\rm L})} c_{n_{\rm L},l,r} |n_{\rm L},l\rangle |N-n_{\rm L},r \rangle,
\label{state}
\eeq
where $|n_{\rm L},l\rangle$ and $|n_{\rm R},r\rangle$ are one-dimer basis states. Given a fixed $n_{\rm L}$ sector of the bi-partite basis, the Hilbert space dimensions of the two subsystems are $d_{\rm L}(n_{\rm L})=n_{\rm L}+1$ and $d_{\rm R}(n_{\rm L})=N-n_{\rm L}+1$. For example, one may use the Fock basis $|n_\alpha,l\rangle=|\ell,m\rangle$ where $\ell=n_\alpha/2$ and $m=(n_{+,\alpha}-n_{-,\alpha})/2=n_\alpha-2n_{-,\alpha}$ with $n_{-,\alpha}=0,1,...n_{\alpha}$. Or, alternatively, the one-dimer energy eigenstates $|n_{\alpha},j_\alpha\rangle$ as in \Eq{upstates}. Regardless of the choice of subsystem basis, the $U(1)$ symmetry of the Hamiltonian in \Eq{Dicke_ham} means that the pure density matrix of the composite system $\hat\rho=|\psi\rangle\langle\psi|$ is block diagonal, with the $n_{\rm L}$ block having dimension $D^{(n_{\rm L})}=d_{\rm L}(n_{\rm L})\times d_{\rm R}(n_{\rm L})$. It is easily verified that $D=\sum_{n_{\rm L}=0}^N D^{(n_{\rm L})}$. Hence, it is possible to study the system's {\em symmetry resolved entanglement spectra}.

\section{Number resolved entanglement}
\label{nrent}

Since $\hat\rho$ is block diagonal, so is the reduced density matrix of the ${\rm L}$ subsystem $\hat\rho_{\rm L}={\rm Tr}_{\rm R} \hat \rho$,
\beq
\hat \rho_{\rm L}=\sum_{n_{\rm L}=0}^N \hat\rho_{\rm L}^{(n_{\rm L})},
\eeq
where the $n_{\rm L}$-th block is,
\beq
\hat \rho_{\rm L}^{(n_{\rm L})}=\sum_{l,l'=1}^{d_{\rm L}(n_{\rm L})} \rho^{n_{\rm L}}_{l,l'} |n_{\rm L},l\rangle\langle n_{\rm L},l'|,\eeq
and its matrix elements are,
\beq
\rho^{n_{\rm L}}_{l,l'}=\sum_{r=1}^{d_{\rm R}(n_{\rm L})} c_{n_{\rm L},l',r}^*c_{n_{\rm L},l,r}~.
\eeq
While the formal dimension of the $n_{\rm L}$ block is $d_{\rm L}(n_{\rm L})$, its rank is $d_{n_{\rm L}}=\min\{d_{\rm L}(n_{\rm L}),d_{\rm R}(n_{\rm L})\}$. Thus, the maximal number of non-zero eigenvalues of $\hat \rho_{\rm L}$ is $d_{\rm L}=\sum_{n_{\rm L}=0}^N d_{n_{\rm L}}$. The reduced density matrix of the $\rm R$ subsystem is identical to $\hat\rho_L$.

\begin{figure}
\centering
\includegraphics[clip=true, width=\columnwidth]{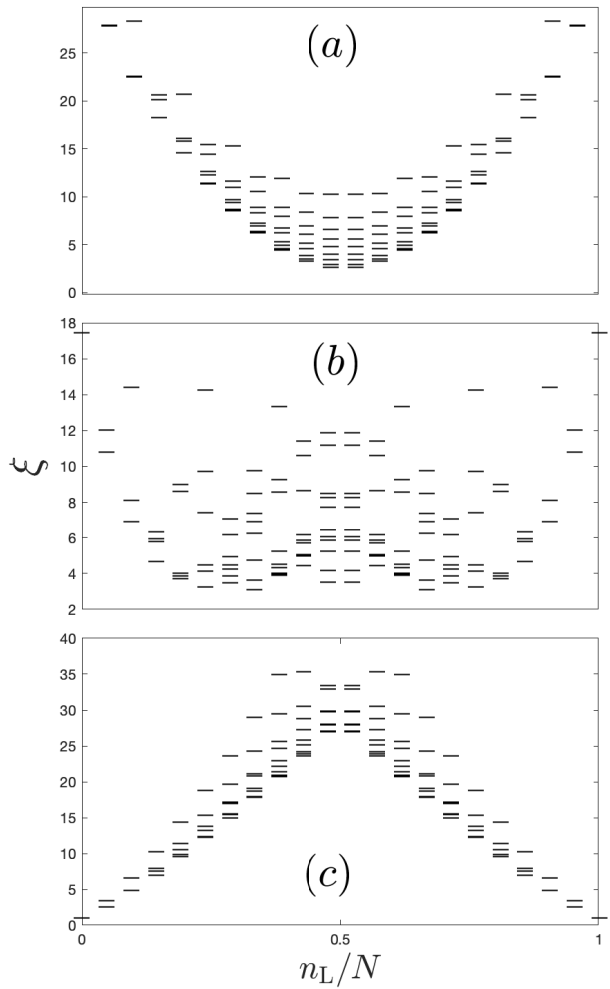}
\caption{{\bf Entanglement spectra:} The number-resolved entanglement spectra of representative states marked by (a) $\circ$, (b) $\triangle$, and (c) $\square$ in Fig.~\ref{ExactSpec}.}
\label{EntSpec}
\end{figure}

Diagonalizing the reduced one-dimer density matrix and expressing the non-zero eigenvalues as 
\beq
\lambda^{(n_{\rm L})}_i=e^{-\xi^{(n_{\rm L})}_i}
\eeq 
with $n_{\rm L}=1,...,N$ and $i=1,...,d_{n_{\rm L}}$, we obtain the symmetry-resolved {\em entanglement spectrum} $\xi^{(n_{\rm L})}_i$ \cite{Li08,Laflorencie16}. The entanglement spectra of the three representative states in \Fig{ExactSpec} are shown in \Fig{EntSpec}. The reduced subsystem density matrix for the island cat states is dominated by few eigenvalues in the populated integrable regions. In particular, for the 'population cat state' in \Fig{ExactSpec}c it is clear that the distribution decays exponentially across the chaos border as befitting particle-tunneling,  whereas for the 'excitation cat state' in \Fig{ExactSpec}d the $n$-distribution is Gaussian, as expected for a superposition of coherent states localized in the two $j$ islands with $n\approx 0$. By contrast, for the chaotic states we observe many eigenvalues of comparable magnitude, spread throughout the chaotic sea. The bi-partite entanglement entropy of the chaotic states should thus be much larger than that of the island-supported eigenstates.

\begin{figure}
\centering
\includegraphics[clip=true, width=\columnwidth]{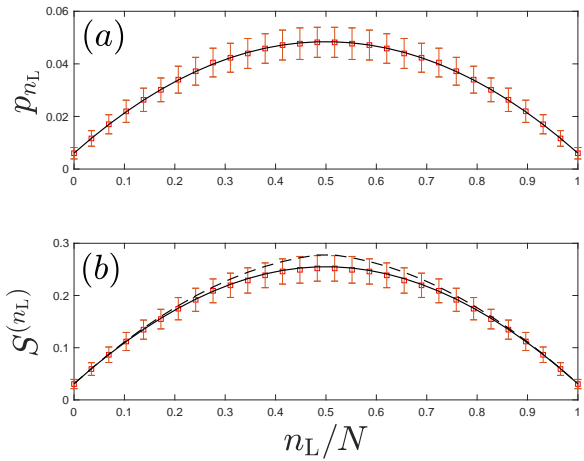}
\caption{{\bf Entanglement of canonical random states:} The particle distribution probability $p_{n_{\rm L}}$ and the number-resolved entanglement entropy $S^{(n_{\rm L})}$ are plotted as a function of $n_{\rm L}$ in panels (a)  and (b), respectively.  Symbols and error bars correspond to the mean and standard deviation over an ensemble of $10^3$ symmetrized canonical random states with $N=29$ particles. The solid line in (a) corresponds to the anticipated ergodic distribution $p_{n_{\rm L}}^{\rm erg}$ of \Eq{pnlerg} whereas the dashed and solid lines in panel (b) correspond to the estimates of $S^{(n_{\rm L})}_{\rm erg}$ in \Eq{Snlerg} and $S^{(n_{\rm L})}_{\rm GOE}$ in \Eq{Snlgoe} , respectively.
}
\label{randomstates}
\end{figure}

Normalizing each block of $\hat \rho_{\rm L}$ according to the the particle-distribution probability,
\beq
p_{n_{\rm L}}= \sum_{i=1}^{d_{\rm n_L}} \lambda^{(n_{\rm L})}_i,
\eeq
one may define the normalized entanglement spectrum $\tilde{\xi}^{(n_{\rm L})}_i=\xi^{(n_{\rm L})}_i+\log p_{n_{\rm L}}$. The total entanglement entropy between the two dimer subsystems,
\beq
S=-{\rm Tr}\left({\hat \rho_{\rm L}}\log{\hat \rho_{\rm L} }\right)=\sum_{n_{\rm L}=0}^N S^{(n_{\rm L})}
\eeq
can thus be written as the sum of the number-resolved entanglement entropies, 
\beq
S^{(n_{\rm L})}=-{\rm Tr}\left({\hat \rho_{\rm L}}^{(n_{\rm L})}\log{\hat \rho_{\rm L}^{(n_{\rm L})}}\right)=-\sum_{i=1}^{d_{\rm n_L}} \lambda^{(n_{\rm L})}_i \log \lambda^{(n_{\rm L})}_i~.
\eeq
The entropies $S^{(n_{\rm L})}$ may be rewritten as,
\beq
S^{(n_{\rm L})}=p_{n_{\rm L}} \tilde S^{(n_{\rm L})}-p_{n_{\rm L}}\log p_{n_{\rm L}}
\eeq
where, \beq
\tilde S^{(n_{\rm L})}=-{\rm Tr}\left(\frac{{\hat \rho_{\rm L}}^{(n_{\rm L})}}{p_{n_{\rm L}}}\log\frac{{\hat \rho_{\rm L}}^{(n_{\rm L})}}{p_{n_{\rm L}}}\right)
\eeq
are the entanglement entropies of the normalized blocks.

\begin{figure}
\centering
\includegraphics[clip=true, width=\columnwidth]{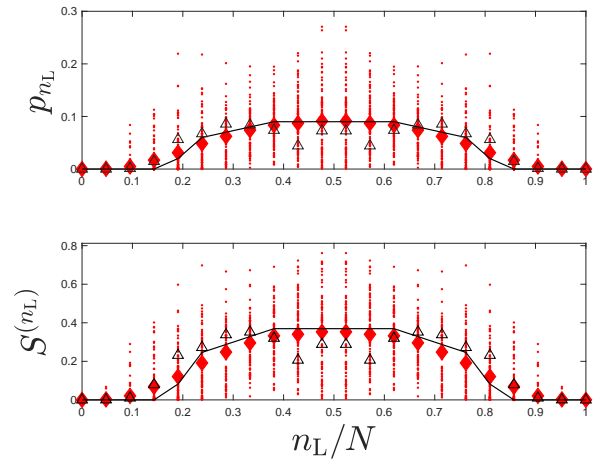}
\caption{{\bf Entanglement of chaos-supported states:} particle distribution probability (a) and number-resolved entanglement entropy (b) for the $D_{\rm ch}^{(J)}=100$ highest ${\cal H}$ states in the $J=11$ shell of the exact spectrum. Diamond symbols mark the mean value, whereas solid lines correspond to the predicted generalized Gibbs ensemble values $p_{J,n_{\rm L}}^{\rm GGE}$ and $S_{\rm GGE}^{(J,n_{\rm L})}$. Triangles mark the number resolved entanglement entropy of the chaotic state of \Fig{ExactSpec}. Parameters are the same as in \Fig{ExactSpec}.}
\label{chaoticstates}
\end{figure}

\section{Entanglement of ergodic states}
Having defined the number resolved entanglement entropy, we turn to predict its expected form for states corresponding to different semiergodic ensembles. These include uniform canonical states, canonical random states that account for fluctuations about the uniform mean, microcanonical states spread on a single energy shell, and GGE states where ergodicity only applies to a restricted fixed $J$ region within the energy shell. 

\label{ergodic}
\subsection{Uniform states}
Consider a completely uniform state $|\psi_{\rm erg}\rangle$ defined by Eq.~(\ref{state}) with  $|c_{n_{\rm L},l,r}|=1/\sqrt{D}$ for all $n_{\rm L},l,r$. The population distribution for such a state is proportional to the density of states 
\beq
p^{\rm erg}_{n_{\rm L}}=\frac{D^{(n_{\rm L})}}{D}~.
\label{pnlerg}
\eeq
The eigenvalues of the reduced one-dimer density matrix are $\lambda_i^{(n_{\rm L)}}=p^{\rm erg}_{n_{\rm L}}/d_{n_{\rm L}}$, hence $\tilde S^{(n_{\rm L})}=\tilde S^{(n_{\rm L})}_{\rm erg}=\log(d_{n_{\rm L}})$ and the number resolved entropy is,
 \beq
S^{(n_{\rm L})}_{\rm erg}=-p^{\rm erg}_{n_{\rm L}} \log\frac{p^{\rm erg}_{n_{\rm L}}}{d_{n_{\rm L}}}
\label{Snlerg}
 \eeq
 The total entanglement entropy is thus,
\beq
S_{\rm erg}=\sum_{n_{\rm L}=0}^N S^{(n_{\rm L})}_{\rm erg}<\log(d_{\rm L})=S_{\rm max}
\label{Serg}
\eeq

\subsection{Canonical random states}
To account for fluctuations over the ergodic mean, consider a random canonical state $|\psi_{\rm GOE}\rangle$ in which, 
\beq
c_{n_{\rm L},l,r} =\frac{z_{n_{\rm L},l,r}}{\sqrt{D}}
\label{psigoe}
\eeq
where $z_{n_{\rm L},l,r}$ are real random numbers, picked from a normal distribution with zero mean and unit variance. Such states emulate the eigenstates of random matrices from a Gaussian Orthogonal Ensemble (GOE), expected for fully chaotic systems that are not restricted to one energy shell (e.g. the kicked rotor). While they are not strictly normalized, the mean of their norm is one and the norm fluctuations rapidly decline with $N$. Thus, we simply renormalize the state vector by the norm ${\cal N}\approx 1$. The mean population distribution remains $p^{\rm erg}_{n_{\rm L}}$ but the expected entanglement entropies $\tilde S^{(n_{\rm L})}$ now include a finite-size fluctuation correction, depending on the partition ratio $\min\{{n_{\rm L}},n_{\rm R}\}/N\le 1/2$ \cite{Page93,Vidmar17}:

\begin{eqnarray}
\tilde S^{(n_{\rm L})}_{\rm GOE}&=&\tilde S^{(n_{\rm L})}_{\rm erg}-\frac{1}{2}\frac{d_{n_{\rm L}}^2}{D_{n_{\rm L}}}\nonumber\\
~&=&\tilde S^{(n_{\rm L})}_{\rm erg}-\frac{1}{2}\frac{\min\{d_{\rm L}(n_{\rm L}),d_{\rm R}(n_{\rm L})\}}{\max\{d_{\rm L}(n_{\rm L}),d_{\rm R}(n_{\rm L})\}}~,
\eeq
so that the expected number-resolved bi-partite entropy for a fully chaotic eigenstate is,
\beq
S^{(n_{\rm L})}_{\rm GOE}=S^{(n_{\rm L})}_{\rm erg}- p_{n_{\rm L}}^{\rm erg} \frac{d_{n_{\rm L}}^2}{2D_{n_{\rm L}}},
\label{Snlgoe}
\eeq
and the total entanglement entropy is,
\beq
S_{\rm GOE}=S_{\rm erg}-\frac{1}{2}\sum_{n_{\rm L}=0}^N  p_{n_{\rm L}}^{\rm erg} \frac{d_{n_{\rm L}}^2}{D_{n_{\rm L}}}~.
\eeq

In \Fig{randomstates} we validate the predictions of \Eq{pnlerg} and \Eq{Snlgoe} by comparison to the mean population distribution and the mean number-resolved entanglement entropy of a numerically generated ensemble of canonical random states. Each random state realisation is symmetrized to the $A_1$ irreducible representation of the $D_2$ group and the mean number distribution and entanglement entropy over all realisations are calculated within each $n_{\rm L}$ sector. The ergodic number distribution and the finite-size fluctuation correction to the entanglement entropy clearly capture the behavior of the canonical random states.

\begin{figure}
\centering
\includegraphics[clip=true, width=\columnwidth]{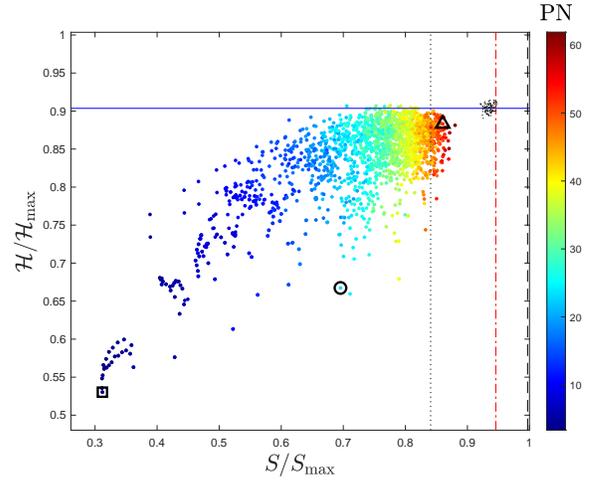}
\caption{{\bf Correlation between chaos and bi-partite entanglement:} The Shanon entropy vs the bi-partite entanglement entropy of the double dimer eigenstates with $N=21$. Symbols point to the marked representative states in Fig.~\ref{ExactSpec}. Black dots correspond to numerically generated canonical random states.  The horizontal solid line marks ${\cal H}_{\rm GOE}$, whereas the vertical lines mark $S_{\rm erg}$ (dashed), $S_{\rm GOE}$ (dash-dotted), and $S_{\rm GGE}\equiv\max_{J}(S_{\rm GGE}^{(J)})$ (dotted).}
\label{ChaosEnt}
\end{figure}

\subsection{GGE states}
Due to the conservation of $E$ and $J$, the ergodicity of the chaos supported eigenstates is incomplete in the sense that they ares restricted to the fixed $J$ region within the energy shell. The expected population distribution for such states will thus differ from $p_{n_{\rm L}}^{\rm GOE}$ and correspond to that of a GGE,
\beq
p_{J,n_{\rm L}}^{\rm GGE}=\frac{D_{\rm ch}^{(J,n_{\rm L})}}{D_{\rm ch}^{(J)}}~,
\label{pgge}
\eeq
where ${D_{\rm ch}^{(J)}}$ is the total number  of unperturbed eigenstates in the chaotic region of the $J$ shell, and $D_{\rm ch}^{(J,n_{\rm L})}$ is the dimension of the fixed $n_{\rm L}$ subset, hence $\sum_{n_{\rm L}} D_{\rm ch}^{(J,n_{\rm L})}={D_{\rm ch}^{(J)}}$. 

The expected number-resolved entanglement entropy for the chaos-supported eigenstates in the coupled-dimers system is accordingly,
\beq
S_{\rm GGE}^{(J,n_{\rm L})}=p_{J,n_{\rm L}}^{\rm GGE}\left(\log D_{\rm ch}^{(J)}-\frac{1}{2}\right),
\label{Sgge}
\eeq
so that the total bi-partite entanglement entropy $S_{\rm GGE}^{(J)}=\log D_{\rm ch}^{(J)}-\frac{1}{2}$ depends on the area of the $J$ shell, rather than on the system's volume. The first term corresponds to the entropy of a uniform state with $p_{J,n_{\rm L}}=1/D_{\rm ch}^{(J)}$, whereas the subtracted factor accounts for maximal fluctuations around this mean value.

In \Fig{chaoticstates}, the expressions for the GGE population-distribution in \Eq{pgge} and entanglement entropy in \Eq{Sgge} are validated by comparison with the corresponding mean values over all the chaotic states in a representative fixed-$J$ shell. While there is an overall good agreement, the entanglement entropy of the chaos supported states is slightly lower than the GGE prediction, indicating larger fluctuations due to the incomplete ergodicity of the eigenstates.

The total bi-partite entanglement entropy $S$ of all the eigenstates of the coupled dimers system is correlated in \Fig{ChaosEnt} with their ergodicity, quantified by the Shanon entropy $\cal H$. The anticipated $S_{\rm GOE}$ and $S_{\rm GGE}$ estimates match the numerical results for canonical random states and chaos-supported eigenstates, respectively. It is also verified that the chaos-supported states are not entirely ergodic, as their Shanon entropy is somewhat below  ${\cal H}_{\rm GOE}$. In comparison, the bipartite entanglement entropy of the island-supported states matches the expectation for macroscopic cat states, for which the reduced density matrix contains only a few (minimally two) nonvanishing eigenvalues.

\section{Conclusions}
\label{conclusions}
The bi-partite entanglement of eigenstates of partitioned systems and its relation to chaotic ergodicity are the subject of a growing body of work \cite{Page93,Deutch10,Santos12,Hamma12,Znidaric07,Alba15,Beugling15,Vidmar17,Garrison18,Murthy19,Huang19,Lydzba20,Lydzba21,Haque22,Molter14,Ares14,Storms14,Deutsch13}. Most effort has so far been concentrated on fermionic systems and spin-chains for which the classical limit is sometimes obscure. The mean-field limit of many-boson system allows for a relatively simple analysis of the classical phase space structure, and a tractable connection to the resulting eigenstate entanglement entropy and its deviations from complete ergodicity. 

The coupled Bose-Josephson system provides an excellent testbed for studying bi-partite entanglement in a mixed phase space with partial ergodicity. We have characterized the global phasespace structure of this system and correlated it with the structure of the U(1) symmetry-resolved entanglement of mid-system eigenstates supported by the different dynamical regions. The dependence of symmetry-resolved entanglement entropy of random states on the relative size of the constituent subsystems was found to follow the Page formula \cite{Page93}. The overall entanglement was found to be restricted by incomplete ergodicity due to the adiabatic invariance of the sum of subsystem actions. Future work will establish how bi-partite entanglement is affected by the breakdown of joson conservation at strong inter-dimer coupling and strong interaction.

\begin{acknowledgments}
This research was supported by the Israel Science Foundation (Grant No.283/18). Valuable discussions with Doron Cohen are greatly appreciated.
\end{acknowledgments}

\end{document}